\documentclass[a4paper]{article}
\usepackage{ISCSLP2026}
\usepackage{ifthen}
\newboolean{blind}
\setboolean{blind}{false} 
\title{A Knowledge-Driven Approach to Target Speech Extraction in the Presence of Background Sound Effects for Cinematic Audio Source Separation (CASS)}
\name{
	\ifthenelse{\boolean{blind}}{Anonymous to ISCSLP}
	{Chun-wei Ho$^1$,Sabato Marco Siniscalchi$^2$, Kai Li$^3$, Chin-Hui Lee$^4$}
}
\address{
  \ifthenelse{\boolean{blind}}{Anonymous to ISCSLP}
  {
  	$^1$ Georgia Institute of Technology, USA \\
    $^2$ University of Palermo, Italy \\
    $^3$ Dolby Laboratory, China
  }
}

\email{
	\ifthenelse{\boolean{blind}}{Anonymous to ISCSLP}
	{chun-wei.ho@gatech.edu}
}

\usepackage{comment}
\usepackage{multirow}
\usepackage{tipa}

\usepackage {tikz}
\usetikzlibrary{shapes,arrows,positioning,calc,matrix, positioning, intersections, arrows, arrows.meta}
\usetikzlibrary{external}
\newcommand{\drawvector}{
    \begin{tikzpicture}
        \foreach \y in {0, 0.9, 1.3}{
            \node[draw, circle, inner sep=3pt] at (0, \y){};
        }
        \node at (0, 0.5){$\vdots$};
    \end{tikzpicture}
}

\begin{document}

\maketitle
\begin{abstract}
  We propose a knowledge-driven approach to speech target extraction in the presence of background sound effects already recorded in cinematic audio. The specific knowledge sources studied are manners of articulation that are detected in speech frames and adopted to form a knowledge vector as a part of features to enhance speech separation and target speech extraction because some short speech segments are often difficult to separate from mixed background sounds. Testing on the recent Sound Demixing Challenge data for cinematic audio source separation (CASS) shows that utilizing articulator-aware knowledge sources produces better separation results than those obtained without using any knowledge, especially for speech segments buried in unspecified background sound events.
\end{abstract}

\section{Introduction}
Cinematic Audio Source Separation (CASS)~\cite{petermann2022cocktail} involves decomposing a movie soundtrack into its constituents: speech, music, and sound effects. Once isolated, these individual ``stems'' facilitate various downstream applications, such as enhancing dialogue, dubbing content into foreign languages, or removing intrusive background noise. Unlike speech or music, cinematic audio is inherently complex and highly variable. Recordings may be captured under near-field or far-field conditions, depending on the scene and filming technique. Sound effects can range from quiet animal noise to intense thunderstorms. In addition, actors might convey a wide spectrum of emotions while performing their lines, which also includes whispering, laughing and shouting~\cite{hasumi2025dnr}. Some movie soundtracks, once captured, are difficult to reproduce or re-record. Therefore, isolating these already-recorded mixed sources through target extraction is becoming a challenging and practical issue in the movie industry. The currently prevailing end-to-end (E2E) modeling strategy often relies on an availability of a large amount of condition-specific training data which is not easily met in CASS.

Conventional source separation~\cite{Makino2007} and target speech extraction techniques~\cite{target-audio-extraction} have been applied to CASS, including Multi-Resolution X-Net (MRX)~\cite{petermann2022cocktail}, band-split RNN~\cite{bsrnn}, Demucs~\cite{defossez2019demucs, HTdemucs}, CodecSep~\cite{banerjee2026codecsep}, and BandIt~\cite{bandit}. These approaches are typically trained in a supervised manner, where the mixture audio serves as the input and the target audio as the output. For source separation, the targets consist of all individual sources within the mixture; whereas, for target audio extraction, the target corresponds to the specific sound source of interest. In this paper, we refer to these approaches as data-driven methods, as those solutions rely solely on learning input-output mappings from the training data without incorporating any auxiliary knowledge or information from the scripts.

Although data-driven methods achieve a reasonable level of performance in terms of signal-to-distortion ratio (SDR) or scale-invariant SDR (SiSDR)~\cite{SDR2006},  a distinctive characteristic of CASS is overlooked, namely the availability of movie scripts. In most separation tasks, such as meeting recordings~\cite{kraaij2005ami}, audio is typically captured directly, often without accompanying transcripts. In contrast, cinematic audio is produced based on pre-written scripts, and actors deliver their lines accordingly. They can provide valuable information to improve the quality of target extraction. For instance, voice activity and overlapping information derived from scripts were used as embedded knowledge vectors to assist the separation process~\cite{ho2026knowledgedrivenapproachmusicsegmentation}. 

In this paper, we adopt a knowledge-driven approach to speech target extraction for cinematic audio in the presence of background sound effects, utilizing the script containing manner-of-articulation cues in speech. Our framework includes aligning cinematic audio with script, extracting manner-of-articulation features from script, and training an articulation-aware separator. Experiments conducted on DNR-nonverbal~\cite{hasumi2025dnr} data for the recent Sound Demixing Challenge~\cite{kim2023sound} demonstrate superior performances using articulation aware embedding to those obtained without utilizing any knowledge source.

\section{Related Work}
\subsection{Speech Source Separation}
A possible technology trend in speech separation, which is a task of decomposing mixed-speech into target and interfering speech, can be seen from a recent review~\cite{Wang2018}. Another trend is target speech extraction, in which the task is to extract only the target speech from a mixture audio as summarized in~\cite{target-audio-extraction}. A deep regression paradigm was introduced for speech enhancement~\cite{Xu2015regression} to map from noisy to clean speech spectrograms. Later it was extended to speech separation by mapping mixed-speech to separated speech of a specific set of one known target speaker and one non-target interfering speaker~\cite{Du2016regression} or two unspecified speakers~\cite{Wang2017gender}.  The technique can also be seen in target speech extraction~\cite{personalized-speech-enhancement}, or sometimes referred to as ``personalized speech enhancement''. 

For speech separation of two speakers, a small training set was enough. In a realistic setting with 5 minutes of training speech from a target speaker plus 20-40 hours of interfering speech, a reasonable quality was observed and separated speech still managed to achieve the highest word accuracy~\cite{Du2016regression} among all competing systems in Speech Separation Challenge (SSC)~\cite{Cooke2010} for automatic speech recognition (ASR)~\cite{Rabiner1993} of mixed-speech. However, to extend the above successes to CASS in the presence of background sound effect, we are faced with a challenge of generating large collections of background sound data in real-world recordings to synthesize a minimally-required amount of mixture training data for learning speech separation or extraction networks.

\subsection{Manner of Articulation}
\label{sec:manner}
Speech attributes, or distinctive speech features~\cite{Fant1973}, are unique units that explain the production of speech sounds by the mouth's articulator and are consistent across all languages, or language universal. Opting for an attribute-based language characterization provides two main benefits over traditional high-level tokens. First, these attributes are universally defined, relieving the need to increase the attribute token set or altering models when introducing new languages. Additionally, these attributes can be updated and improved as new linguistic data is collected, regardless of the languages. Previous works on attribute modeling focused on automatic speech attribute transcription (ASAT)~\cite{Lee2004,Lee2007,Lee2013}, which built and utilized various attribute detectors with successes in several speech applications~\cite{Lee2013,Siniscalchi2012,Chen2014,Li2016}.

In acoustic phonetics (e.g., ~\cite{Fant1973}), manner of articulation describes how airflow is shaped or restricted within the vocal tract to produce a particular speech sound. The IPA (International Phonetic Alphabet~\cite{IPA} classifies the manners into the six categories, namely: nasal (NAS, e.g., /\textipa{m}/, /\textipa{n}/), approximant (APR, e.g., /\textipa{l}/, /\textipa{w}/, /\textipa{j}/), flap (FLP, e.g., /\textipa{2}/ in the middle vowel for ''butter''), plosive or stop (STP, e.g., /\textipa{p}/, /\textipa{k}/, /\textipa{d}/), fricative (FRC, e.g., /\textipa{s}/, /\textipa{f}/, /\textipa{z}/) and affricate (AFR, e.g., /\textipa{tS}/).

In addition to the aforementioned consonant categories, vowels (e.g., /\textipa{a}/, /\textipa{i}/, /\textipa{u}/) represent another important class of speech sounds, typically characterized by the articulator positions in the oral cavity, such as tongue height, tongue backness, and lip rounding, which directly determines the formant frequencies. Accordingly, some approaches~\cite{peterson1952control, hillenbrand1995acoustic} define vowels in terms of their formant frequency patterns, or based on places of articulation~\cite{Fant1973, Lee2013}.

Understanding manner of articulation is important for cinematic audio processing, as certain speech categories can easily be confused with other common sounds in cinematic audio. For example, fricatives may be mistaken for white noise or "shh" effects in music, while plosives can be confused with percussive elements, such as a snare drum hit.

\subsection{Knowledge-driven Separation}
In contrast to conventional data-driven approaches, knowledge-driven separation leverages upon prior knowledge sources, such as instrument categories or voice activity (VA) boundaries, as auxiliary cues to facilitate source separation. For example, in~\cite{ho2026knowledgedrivenapproachmusicsegmentation}, two strategies were proposed to incorporate such knowledge information: (1) VA is used to guide training data selection and also utilized as auxiliary inputs to the separation model; and (2) scores are used to guide music segmentation. They were empirically shown to be effective for music separation.

Specifically, using knowledge to select training data helps the model identify cleaner segments, leading to more stable training. In contrast, incorporating knowledge as auxiliary input provides additional guidance, enabling the model to better shape the output audio during both training and inference.

\section{Proposed Speech Separation Framework}
\label{sec: Proposed}
\subsection{Audio Alignment with Movie Script}
\label{sec:3.1}
The cinematic audio often comes with speech transcription along with the start and end times of each line. However, it does not include the manner-of-articulation categories or their corresponding timestamps. Therefore, to effectively incorporate manner-of-articulation information into the separation process, it is necessary to first detect the manner of articulation and estimate its start and end times.

A straightforward approach is to perform ``forced-alignment'' between the transcription and the audio. The transcription must first be tokenized into smaller units, referred to as tokens, which may be words, subwords, phonemes, or attributes. Assuming the token length of a line is $L$, we denote the start time (in terms of frames), end time, and token ID of the $i$-th token as $s_i$, $e_i$, and $t_i$, respectively, where $i \in \{1, \dots, L\}$. $t_i$ can be directly inferred from the transcription and doesn't require any estimation.

Given a recorded line, consisting of $N$ frames, the corresponding acoustic features are denoted by $x_k \in \mathbb{R}^D$, where $k \in \{1, \dots, N\}$ and $D$ denote the feature vector dimensions. An estimator is trained to approximate the posterior probability $\hat{P}(x_{1 : N} \mid s_{1 : L}, e_{1 : L})$. The estimator may adopt any model architecture and training strategy. The objective is to maximize this posterior probability.
\begin{equation}
    \begin{aligned}
    (\hat{s}_{1:L}, \hat{e}_{1:L}&)=\arg\max
    \hat{P}\left(x_{1:N} \mid s_{1:L}, e_{1:L}\right),\\
    &\text{where}\quad 1 \leq s_1 < e_1 < \cdots < s_L < e_L \leq N
    \end{aligned}
\end{equation}

For each script line, we estimate a distinct set of $\hat{s}_i$ and $\hat{e}_i$. The predicted start and end times of each token in the transcription are obtained by adding $\hat{s}_i$ and $\hat{e}_i$ to the corresponding line-level start time.

\subsection{Articulation-aware Separator}
 
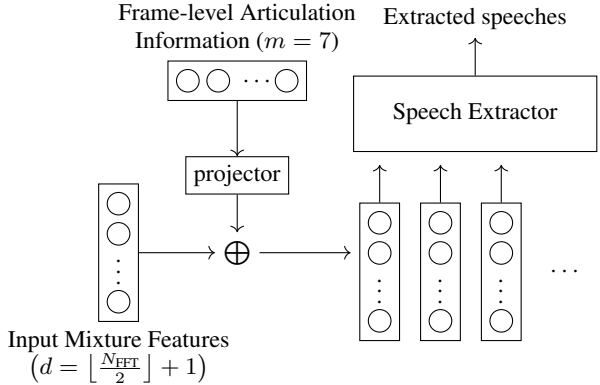
\begin{figure}
    \centering
    \begin{tikzpicture}
    \node[draw, label={[align=center]below:Input Mixture Features\\ $\left(d=\left\lfloor\frac{N_\text{FFT}}{2}\right\rfloor+1\right)$}](input frame){\drawvector};
            
            \node[right=of input frame](sum){$\bigoplus$};
            \node[draw, above=.5 of sum](projector){projector};
            \node[draw, label={[align=center]east:Frame-level Articulation\\Information ($m=7$)}, above=of projector, rotate=90, anchor=center](knowledge){\drawvector};
    
            \node[right=1.2 of sum](separator input){
                \begin{tikzpicture}
                    \foreach \x in {0, 0.8, ..., 1.7}{
                        \node[draw](a) at (\x, 0){\drawvector};
                        \draw[->] (a.north) -- ($(a.north)+(0,0.5)$);
                    }
                    \node at (2.4, 0){$\cdots$};
                \end{tikzpicture}
            };
            \node[above=.05 of separator input, draw, minimum width=3.2cm, minimum height=1cm](separator){Speech Extractor};
    
            \node[above=.5 of separator](output){Extracted speeches};
            
            \draw[->] (knowledge) -- (projector);
            \draw[->] (projector) -- (sum);
            \draw[->] (input frame) -- (sum);
            \draw[->] (sum) -- (separator input);
            \draw[->] (separator) -- (output);
    \end{tikzpicture}
    \caption{An illustration of an articulation-aware speech separator. Frame-level speech attribute (nasal, approximant, flap, plosive, fricative, affricate, and vowels) is provided as auxiliary input to the system. Speech extractor can be implemented using any reasonable target audio extraction mechanism.}
    \label{fig:knowledge-aware separator}
\end{figure}

Figure~\ref{fig:knowledge-aware separator} shows the overall architecture of the articulation-aware separator. We first convert sentence labels into phoneme sequences using the Speech Assessment Methods Phonetic Alphabet (SAMPA). Each consonant is then mapped into its corresponding manner of articulation based on the six defined categories in Section ~\ref{sec:manner}. together a special token, ``vwl'', for all vowels, each frame is represented by a $m$-dim articulation vector ($m=7$, shown in the upper part of Figure~\ref{fig:knowledge-aware separator}) where each dimension indicates the presence or absence of a specific attribute label at the current time frame. Thus, the articulation vectors are  aligned with the audio features at the same frame resolution.  These frame-level features are then passed through a frame-wise linear projection layer (an $m$-by-$d$ projector) to match the dimensionality of the extracted audio features. 

In this study, for simplicity, the input to the speech extractor is represented by a sum of the projected articulation features and the audio features (with dimension $d$ as shown in the lower left part of Figure~\ref{fig:knowledge-aware separator}). Due to the mismatch in size here between $m$ (7) and $d$ (1025), other features, such as attention mechanisms~\cite{Subakan2021} and latent variable embedding~\cite{Hu2025} will be explored in the future. As for the speech extractor or separator, it can be in any architecture form that takes the combined features as input and outputs the speech stems from cinematic data.

\section{Experiments and Result Analysis}

All experiments were conducted on DNR-non-vertical~\cite{hasumi2025dnr}. It consists of 1,000 mixture recordings for training, 50 for validation, and 100 for testing, with each utterance exactly 1-minute long, containing a mixture of speech, music, and sound effects.

In this work, {\em target speech} is defined as the combination of reading voice and non-verbal human vocalizations. In DNR-Nonverbal, the reading voice stems are taken from LibriSpeech~\cite{librispeech} and thus containing the transcriptions, while non-verbal human sounds, such as laughter, whispering, crying, sobbing, screaming, sighing, and shouting, are drawn from FSD50K~\cite{fonseca2021fsd50k}. The {\em interfering components} include music and sound effects, with music samples taken from FMA~\cite{defferrard2016fma}, and sound effects (e.g., vehicle noises, animal sounds, and thunder) obtained from FSD50K~\cite{fonseca2021fsd50k}. The {\em mixture audio} is defined as the mixture of target speech and interfering components.

The data set also comes with scripts, providing the start and end times of each line along with the exact transcription. It also contains information about the instrument types in the music and the categories of the sound effects, although these details were not used as input to the model in our experiments.

\begin{figure}
    \centering
    \includegraphics[width=\linewidth]{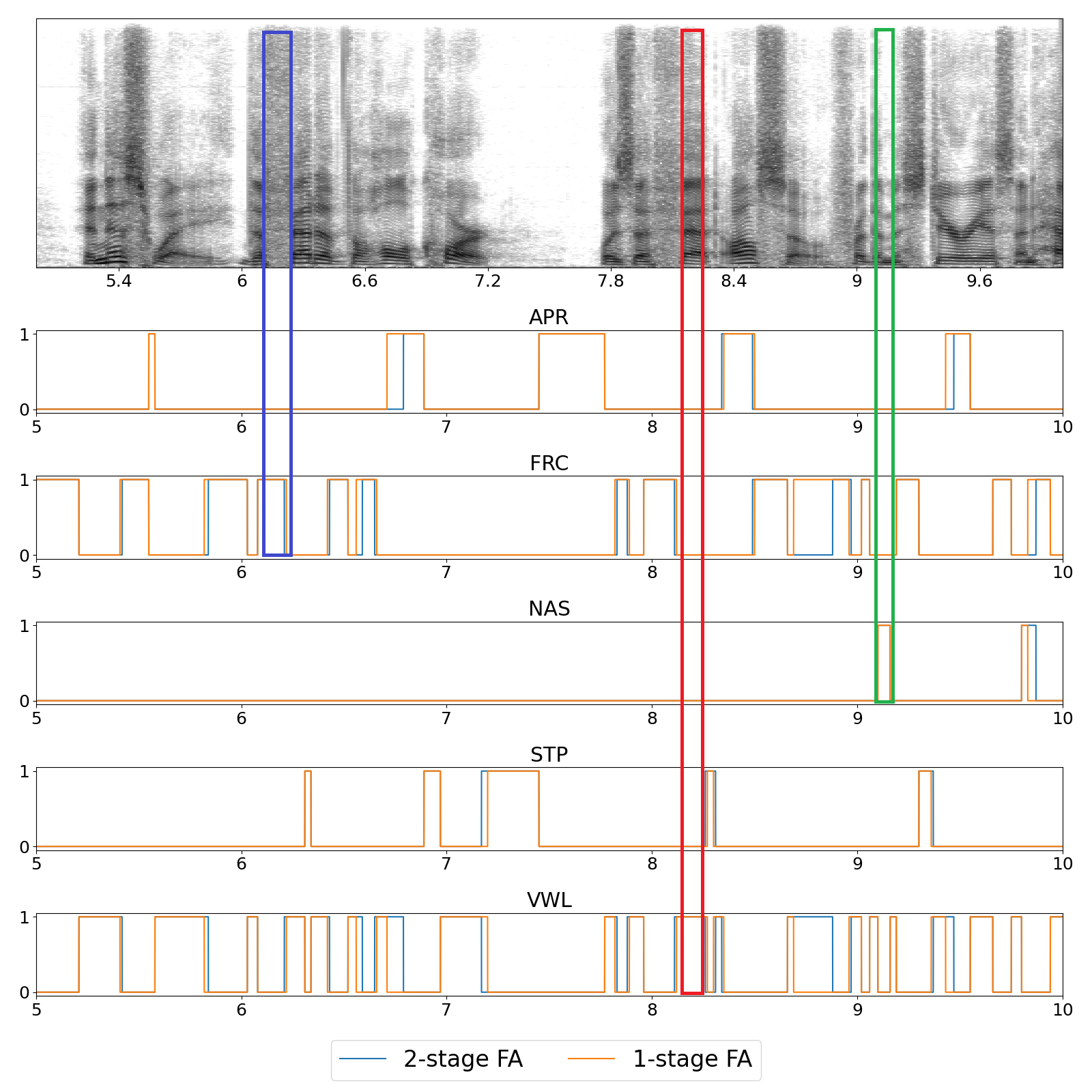}
    \caption{A visualization of the aligned articulation labels on the test set under both 1-stage and 2-stage forced alignments.}
    \label{fig:visualization}
\end{figure}

\subsection{Aligning Transcript with Mixture Audio}
\subsubsection{Experimental Setup}
\label{sec:sec4.2.1}
Data segmentation is an important step in our study because not all timestamps contain speech. To this end, mixture segments, containing reading voice, were cropped according to the transcript. Each segment is paired with the corresponding line and its transcription, and then tokenized into manner-of-articulation units (as described below) for alignment.


The acoustic features discussed in Section \ref{sec:3.1}  are based on the Mel-frequency cepstral coefficients (MFCCs)~\cite{mfcc}. Specifically,  39-dimensional MFCC vectors are extracted from the audio using a window analysis of 25 ms and a hop size of 10 ms. To find the timestamps of manner sequences, we follow the classical Viterbi algorithm~\cite{Rabiner1993} to segment continuous speech into phonemes or attributes. Hidden Markov models (HMMs)~\cite{Rabiner1989}, capable of capturing both spectral and temporal variations in speech, are adopted as models to perform forced alignment (FA)~\cite{Rabiner1989} to obtain unit boundaries given unit transcriptions. 

Two different alignment strategies can be used: (a) 1-stage FA: train GMM-HMMs (Model 1) using the target audio from the training set and then perform forced alignment on the mixture audio in the test set; and
(b) 2-stage FA (Practical Alignment): use GMM-HMMs obtained in 1-stage FA to perform forced alignment on the mixture audio in the test set to estimate both unit sequences and boundaries. Next, train a second set of GMM-HMMs (Model 2). This process can be repeated to obtain refined boundaries.

In this study, we model each manner class with a 5-state GMM-HMM~\cite{Rabiner1993}, trained using the Hidden Markov Toolkit (HTK)~\cite{young1999htk}. The resulting alignments are then used for training in the next stage for speech separation and mixture target extraction. This is similar to a recently emerged psuedo-labelling or self-training~\cite{SelfTraining} strategy in deep learning for enriching labeled data sets and retraining classification models. We believe that 2-stage FA will produce better models than Model 1 obtained with 1-stage FA in real-world scenarios for cinematic audio.

\begin{table*}[!ht]
    \centering
    \caption{Speech SDR in different articulation  categories. The percentages in the train and test sets are about the same.}
    \label{tab:categorized-SDR}
    \begin{tabular}{p{.6\columnwidth} | c c c c c c c}
        \toprule
        \multirow{2}{*}{Method} & \multicolumn{7}{c}{SDR (dB)}\\\cline{2-8}
        & AFR & APR & FLP & FRC & NAS & STP & VWL \\
        \midrule
        Percentage in the train/test set (\%) & 1.12 & 9.00 & 5.89 & 18.31 & 13.62 & 22.66 & 29.40 \\
        \hline
        BandIt baseline & \textbf{14.26} & 16.46 & \textbf{15.48} & \textbf{13.55} & 13.93 & 13.70 & 15.96 \\
        BandIt + VA + 2-stage FA & 13.44 & \textbf{16.59} & 14.94 & \textbf{13.55} & \textbf{14.18} & \textbf{13.78} & \textbf{16.25} \\
        \bottomrule
    \end{tabular}
\end{table*}

\begin{table}[]
    \centering
    \caption{Overall speech extraction performance on the DNR-nonverbal data set. VA stands for voice activities.}
    \begin{tabular}{p{.55\columnwidth} | c c}
        \toprule
        \multirow{2}{*}{Method} & \multicolumn{2}{c}{Speech Performance} \\\cline{2-3}
        & SDR (dB) & SiSDR (dB)  \\
        \midrule
        BSRNN~\cite{hasumi2025dnr} & 9.30 & - \\
        SepReformer~\cite{ho2026knowledgedrivenapproachmusicsegmentation} & 7.68 & - \\
        SepReformer + VA~\cite{ho2026knowledgedrivenapproachmusicsegmentation} & 11.03 & - \\
        \hline
        BandIt baseline & 12.01 & 11.26 \\
        BandIt + VA & 12.12 & 10.78 \\
        BandIt + VA + 1-stage FA & 12.97 & 12.36 \\
        BandIt + VA + 2-stage FA & \textbf{13.01} & \textbf{12.43} \\ 
        \bottomrule
    \end{tabular}
    \label{tab:overall}
\end{table}

\subsubsection{Results and Performance Analyses}
A sample of the alignment results from a test utterance is visualized in Figure~\ref{fig:visualization}. As shown, the Practical Alignment (in orange lines obtained with 2-stage FA proposed in Section ~\ref{sec:sec4.2.1}) behaves similarly to the oracle alignment (in blue lines), indicating that even with mixture audio, 2-stage FA produces alignments that are reasonably consistent with the oracle boundaries.

From Figure~\ref{fig:visualization}, we can also observe several spectral characteristics. For example, vowel sounds (an example shown in the red box) typically exhibit clear formant frequencies in the spectrogram, while fricatives (an example shown in the blue box) contain stronger high-frequency components. Both alignments capture these properties quite well. Nasal sounds (an example shown in the green box) resemble vowel regions but appear whiter. Clearly, Practical Alignment still produces reasonable results, although with occasional false negatives. This is likely because nasal sounds generally have lower energy and are more susceptible to perturbations in the mixture audio with background sound effects.



\subsection{Articulation-aware Speech Target Extraction}
\subsubsection{Experimental Setup}
During training of the articulation-aware separator, the audio is cropped into 6-second segments. The cropping process is not entirely random; it is guided by the voice activity information from the script. Specifically, there is a 25\% probability that a training chunk is directly sampled from an original mixture in the training set that contains non-silent speech.

For the remaining 75\%, the data is generated through random mixing. In this case, a 6-second non-silent speech segment from utterance A is mixed with a 6-second music segment from utterance B and a 6-second sound effect segment from utterance C, where utterances A-B-C are randomly sampled from the train set. The selected speech segments are always non-silent, while there is a 20\% probability of excluding music and a 20\% probability of excluding sound effects. Each stem is randomly scaled by a linear factor between 0.7 and 1.3 before mixing.

The audio was sampled at 44.1 kHz and processed using a 2048-point short-time Fourier transform (STFT) with a window size of 2048 and a hop size of 300. The extracted features ($d=1025$) were then processed as illustrated in Figure~\ref{fig:knowledge-aware separator}, with the speech extractor implemented using the BandIt model~\cite{bandit}. During inference, all mixture segments containing non-silent speech are collected and processed. Each segment is processed using an overlap-add strategy for reconstruction.

\subsubsection{Results and Performance Analyses}
First, we analyze SDR~\cite{SDR2006} results across different articulation categories. For each test utterance, we select a specific articulation label and mute all segments corresponding to other articulation labels or those that do not belong to human speech. We then compute the SDR value over each entire utterance. The results are summarized for the 7 articulation categories in Table~\ref{tab:categorized-SDR}.

As clearly indicated, improvements are observed in most categories, except for affricate (AFR) and Flap (FLP). However, since only 1.12\% and 5.89\% of the data fall into these two categories, the average performance still slightly improves. The degraded SDR results in these two categories may be due to insufficient training data. This suggests that collecting more data for these underrepresented categories could further improve the proposed technique. As for the category ''VWL'' in the rightmost column in Table~\ref{tab:categorized-SDR}, with the most training data, the improvement is the most significant (from 15.96 dB to 16.25 dB).

Next, the overall SDR performance on the DNR-nonverbal data set is summarized in Table~\ref{tab:overall}. In the three rows in the upper block, we show state-of-the-art performances from previous studies. As in the three rows in the lower block, The baseline result of 12.01 dB SDR is obtained with the BandIt model trained and tested without utilizing any knowledge source. Adding the VAD information (results in the second row) provides only a slight improvement to 12.12 dB SDR. This may be because BandIt is a well-designed architecture for cinematic audio data and therefore the BandIt model augmented with VAD and alignment information manages to achieve the state-of-the-art performance on the mixture speech stems. 

Nonetheless, incorporating alignment information into articulation-aware separator (bottom row) almost increases by 1 dB (SDR from 12.12 dB to 13.01 dB). This suggests that the gain comes mainly from the proposed knowledge-driven framework rather than the VA cues alone. As expected, the proposed 2-stage FA in Section ~\ref{sec:sec4.2.1} produces better models (Model 2) than Model 1 obtained with 1-stage FA because Model 2 generates better alignments that fit the mixture cinematic data well at the articulation level (SDR at 13.01 dB slightly better than 12.97 dB obtained with Model 1 shown in the row above the bottom row). We also expect that the larger the disparity between training and testing data, the more the SDR and SiSDR improvements from 1-stage to 2-stage FA.



\section{Summary and Future Work}
In this paper, we propose an articulation-aware speech separation and target extraction framework for mixture data in the presence of background sound effects in cinematic audio source separation (CASS). The algorithm consists of two steps: articulation-based alignment and articulation-aware separation. Experimental results demonstrate that articulation labels can be reliably aligned to mixture audio and provide useful information for the subsequent separation stage. By incorporating articulation information, the overall speech separation performance on cinematic audio is improved. Larger gains are observed in articulation categories that occur more frequently in movie audio.

In future work, We will utilize other knowledge sources in movie, music and speech, such as speech acts, music scores, and places of articulation~\cite{Lee2013}. Various features, attention networks and deep learning architectures for designing speech separators and target extractors will be studied. Due to a scarcity of training data in mixed-speech embedded in background sounds, adaptation and fine-tuning techniques, e.g., model finetuning and adaptive learning, applied to attention mechanisms~\cite{Subakan2021} and latent variables~\cite{Hu2025}, will also be explored.
\pagebreak
\bibliographystyle{IEEEtran}
\bibliography{mybib}

@ARTICLE{bandit,
  author={Watcharasupat, Karn N. and Wu, Chih-Wei and Ding, Yiwei and Orife, Iroro and Hipple, Aaron J. and Williams, Phillip A. and Kramer, Scott and Lerch, Alexander and Wolcott, William},
  journal={IEEE Open Journal of Signal Proc.},
  title={A Generalized Bandsplit Neural Network for Cinematic Audio Source Separation},
  year={2024},
  volume={},
  number={},
  pages={73-81},
  doi={10.1109/OJSP.2023.3339428}}

@article{hasumi2025dnr,
  title={DnR-nonverbal: Cinematic Audio Source Separation Dataset Containing Non-Verbal Sounds},
  author={Hasumi, Takuya and Fujita, Yusuke},
  journal={arXiv preprint arXiv:2506.02499},
  year={2025}
}

@inproceedings{petermann2022cocktail,
  title={The cocktail fork problem: Three-stem audio separation for real-world soundtracks},
  author={Petermann, Darius and Wichern, Gordon and Wang, Zhong-Qiu and Le Roux, Jonathan},
  booktitle={ICASSP 2022-2022 IEEE International Conference on Acoustics, Speech and Signal Proc. (ICASSP)},
  pages={526--530},
  year={2022},
  organization={IEEE}
}

@misc{
banerjee2026codecsep,
title={CodecSep: Prompt-Driven Universal Sound Separation on Neural Audio Codec Latents},
author={Adhiraj Banerjee and Vipul Arora},
year={2026},
url={https://openreview.net/forum?id=MDHVDfUrDz}
}

@ARTICLE{target-audio-extraction,
  author={Zmolikova, Katerina and Delcroix, Marc and Ochiai, Tsubasa and Kinoshita, Keisuke and Černocký, Jan and Yu, Dong},
  journal={IEEE Signal Processing Magazine}, 
  title={Neural Target Speech Extraction: An overview}, 
  year={2023},
  volume={40},
  number={},
  pages={8-29},
  doi={10.1109/MSP.2023.3240008}
  }

@article{defossez2019demucs,
  title={Demucs: Deep extractor for music sources with extra unlabeled data remixed},
  author={D{\'e}fossez, Alexandre and Usunier, Nicolas and Bottou, L{\'e}on and Bach, Francis},
  journal={arXiv preprint arXiv:1909.01174},
  year={2019}
}

@INPROCEEDINGS{HTdemucs,
  author={Rouard, Simon and Massa, Francisco and Défossez, Alexandre},
  booktitle={IEEE International Conference on Acoustics, Speech and Signal Proc. (ICASSP)},
  title={Hybrid Transformers for Music Source Separation},
  year={2023},
  volume={},
  number={},
  pages={1-5},
  doi={10.1109/ICASSP49357.2023.10096956}}

@ARTICLE{bsrnn,
  author={Luo, Yi and Yu, Jianwei},
  journal={IEEE/ACM Transactions on Audio, Speech, and Language Proc.},
  title={Music Source Separation With Band-Split RNN},
  year={2023},
  volume={31},
  number={},
  pages={1893-1901},
  doi={10.1109/TASLP.2023.3271145}}

@inproceedings{kraaij2005ami,
  title={The AMI meeting corpus},
  author={Kraaij, Wessel and Hain, Thomas and Lincoln, Mike and Post, Wilfried},
  booktitle={Proc. International Conference on Methods and Techniques in Behavioral Research},
  pages={1--4},
  year={2005}
}

@article{IPA,
 ISSN = {00978507, 15350665},
 URL = {http://www.jstor.org/stable/414611},
 author = {Peter Ladefoged},
 journal = {Language},
 number = {},
 pages = {550--552},
 publisher = {Linguistic Society of America},
 title = {The Revised International Phonetic Alphabet},
 urldate = {2026-02-18},
 volume = {66},
 year = {1990}
}

@article{peterson1952control,
  title={Control methods used in a study of the vowels},
  author={Peterson, Gordon E and Barney, Harold L},
  journal={The Journal of the acoustical society of America},
  volume={24},
  number={2},
  pages={175--184},
  year={1952},
  publisher={Acoustical Society of America}
}

@article{hillenbrand1995acoustic,
  title={Acoustic characteristics of American English vowels},
  author={Hillenbrand, James and Getty, Laura A and Clark, Michael J and Wheeler, Kimberlee},
  journal={The Journal of the Acoustical society of America},
  volume={97},
  number={},
  pages={3099--3111},
  year={1995},
  publisher={Acoustical Society of America}
}

@article{defferrard2016fma,
  title={FMA: A dataset for music analysis},
  author={Defferrard, Micha{\"e}l and Benzi, Kirell and Vandergheynst, Pierre and Bresson, Xavier},
  journal={arXiv preprint arXiv:1612.01840},
  year={2016}
}

@article{fonseca2021fsd50k,
  title={Fsd50k: an open dataset of human-labeled sound events},
  author={Fonseca, Eduardo and Favory, Xavier and Pons, Jordi and Font, Frederic and Serra, Xavier},
  journal={IEEE/ACM Transactions on Audio, Speech, and Language Proc.},
  volume={30},
  pages={829--852},
  year={2021},
  publisher={IEEE}
}

@INPROCEEDINGS{librispeech,
  author={Panayotov, Vassil and Chen, Guoguo and Povey, Daniel and Khudanpur, Sanjeev},
  booktitle={2015 IEEE International Conference on Acoustics, Speech and Signal Proc. (ICASSP)},
  title={Librispeech: An ASR corpus based on public domain audio books},
  year={2015},
  volume={},
  number={},
  pages={5206-5210},
  doi={10.1109/ICASSP.2015.7178964}}

@misc{young1999htk,
	title        = {The HTK book},
	author       = {S. Young and G. Evermann and M. Gales and T. Hain and D. Kershaw and X. Liu and G. Moore and J. Odell and D. Ollason and D. Povey and V. Valtchev and P. Woodland},
	year         = {1999},
	publisher    = {version}
}

@article{Wang2018,
  title={Supervised speech separation based on deep learning: an overview},
  author={Wang, D. L. and Chen, J.},
  journal={{IEEE/ACM} Transactions on Audio, Speech, and Language Proc.},
  volume={26},
  number={},
  pages={1702--1726},
  year={2018}
  }

@article{SDR2006,
  title={Performance measurement in blind audio source separation},
  author={Vincent, E. and Gribonval, R. and Fevotte C.},
  journal={IEEE Transactions on Audio, Speech, and Language Proc.},
  volume={14},
  number={},
  pages={1462--1469},
  year={2006}
  }

@article{Xu2015regression,
  title={A regression approach to speech enhancement based on deep neural networks},
  author={Xu, Yong and Du, Jun and Dai, Li-Rong and Lee, Chin-Hui},
  journal={IEEE/ACM Transactions on Audio, Speech, and Language Proc.},
  volume={23},
  pages={7--19},
  year={2015},
  publisher={IEEE}
}

@article{Du2016regression,
  title={A regression approach to single-channel speech separation via high-resolution deep neural networks},
  author={Du, Jun and Tu, Yanhui and Dai, Li-Rong and Lee, Chin-Hui},
  journal={IEEE/ACM Transactions on Audio, Speech, and Language Proc.},
  volume={24},
  number={},
  pages={1424--1437},
  year={2016},
  publisher={IEEE}
}

@article{Wang2017gender,
  title={A gender mixture detection approach to unsupervised single-channel speech separation based on deep neural networks},
  author={Wang, Yannan and Du, Jun and Dai, Li-Rong and Lee, Chin-Hui},
  journal={IEEE/ACM Transactions on Audio, Speech, and Language Proc.},
  volume={25},
  number={},
  pages={1535--1546},
  year={2017},
  publisher={IEEE}
}

@article{mfcc,
  title={Comparison of parametric representation for monosyllabic word recognition in continuous spoken utterances},
  author={Davis, S. and Mermestein, P.},
  journal={IEEE transactions on acoustics, speech, and signal proc.},
  volume={28},
  number={},
  pages={357--366},
  year={2000},
  publisher={IEEE}
}

@book{Rabiner1993,
author = {Rabiner, Lawrence and Juang, Biing-Hwang},
title = {Fundamentals of speech recognition},
year = {1993},
isbn = {0130151572},
publisher = {Prentice-Hall, Inc.},
address = {USA}
}

@inproceedings{Rabiner1989,
  title={A tutorial on hidden Markov Models and selected applications in speech Recognition},
  author={Rabiner, L. R.},
  booktitle={Proceedings of the IEEE},
  volume={77},
  number={},
  pages={257--286},
  year={1989}
  }

@book{Fant1973,
  title={Speech Sounds and Features},
  author={Fant, G.},
  year = {1973},
  isbn = {0130151572},
  publisher = {MIT Press},
  address = {USA}
  }

@article{Hu2025,
  title={Variational Bayesian Learning for Deep Latent Variables for Acoustic Knowledge Transfer},
  author={Hu, Hu and Siniscalchi, Marco and Yang, Chao-Han and Lee, Chin-Hui},
  journal={IEEE/ACM Transactions on Audio, Speech, and Language Proc.},
  volume={33},
  number={},
  pages={719--730},
  year={2025},
  publisher={IEEE}
}

@inproceedings{Subakan2021,
  title={Attention is all you need in speech separation},
  author={Subakan, Cem and Ravanelli, Marco and Cornell, Samuele and Bronzi, Mirko and Zhong, Jianyuan},
  booktitle={{ICASSP}},
  pages={261--265},
  year={2021}
  }

@article{Cooke2010,
  title={Monoral speech separation and recognition challenge},
  author={Cooke, M. and Hersey, J. R. and Rennie, S. J.},
  journal={Computer Speech and Language},
  volume={24},
  number={},
  pages={1--15},
  year={2010},
  publisher={Elsevier}
}

@article{kim2023sound,
	title        = {Sound Demixing Challenge 2023 Music Demixing Track Technical Report: TFC-TDF-UNet v3},
	author       = {Kim, Minseok and Lee, Jun Hyung and Jung, Soonyoung},
	year         = {2023},
	journal      = {arXiv preprint:2306.09382}
}

@article{Chen2014,
  author={Chen, I-Fan and Siniscalchi, Sabato Marco and Lee, Chin-Hui},
  journal={IEEE International Conference on Acoustics, Speech and Signal Proc. (ICASSP)},
  title={Attribute based lattice rescoring in spontaneous speech recognition},
  year={2014}
}

@article{Li2016,
  title={Improving non-native mispronunciation detection and enriching diagnostic feedback with DNN-based speech attribute modeling},
  author={Wei Li and Sabato Marco Siniscalchi and Nancy F. Chen and Chin-Hui Lee},
  journal={IEEE International Conference on Acoustics, Speech and Signal Proc. (ICASSP)},
  year={2016}
}

@ARTICLE{Lee2013,
  author={Lee, Chin-Hui and Siniscalchi, Sabato Marco},
  journal={Proc. of the IEEE},
  title={An Information-Extraction Approach to Speech Processing: Analysis, Detection, Verification, and Recognition},
  year={2013},
  volume={101},
  number={},
  pages={1089-1115}
}

@ARTICLE{Siniscalchi2012,
  author={Siniscalchi, Sabato Marco and Lyu, Dau-Cheng and Svendsen, Torbjørn and Lee, Chin-Hui},
  journal={IEEE Transactions on Audio, Speech, and Language Proc.},
  title={Experiments on Cross-Language Attribute Detection and Phone Recognition With Minimal Target-Specific Training Data},
  year={2012},
  volume={20},
  number={},
  pages={875-887},
}

@article{Lee2004,
  title={From knowledge-ignorant to knowledge-rich modeling: A new speech research paradigm for next generation automatic speech recognition},
  author={Lee, Chin-Hui},
  journal={International Conference on Spoken Language Proc. (ICSLP)},
  year={2004}
}

@article{Lee2007,
  title={An overview on automatic speech attribute transcription (ASAT)},
  author={Chin-Hui Lee and Mark A. Clements and Sorin Dusan and Eric Fosler-Lussier and Keith Johnson and Biing-Hwang Juang and Lawrence R. Rabiner},
  journal={Interspeech},
  year={2007}
}

@article{SelfTraining,
  title={Self-training: a survey},
  author={M.-R. Amini and V. Felfanov and L. Pauletto and L. Hadjadj and E. Devijver and Y. Maximov},
  journal={Neurocomputing},
  year={2025},
  publisher={Elsevier}
}

@article{personalized-speech-enhancement,
  author={Eskimez, Sefik Emre and Yoshioka, Takuya and Wang, Huaming and Wang, Xiaofei and Chen, Zhuo and Huang, Xuedong},
  journal={IEEE International Conference on Acoustics, Speech and Signal Proc. (ICASSP)}, 
  title={Personalized speech enhancement: new models and Comprehensive evaluation}, 
  year={2022},
  volume={},
  number={},
  pages={356-360},
  doi={10.1109/ICASSP43922.2022.9746962}
  }

@book{Makino2007,
author = {Makino, S. and Lee, T.-J. and Sawada, H.},
 title = {Blind Speech Separation},
 year = {2007},
 isbn = {0792393961},
 publisher = {Springer},
 address = {USA},
}

@misc{ho2026knowledgedrivenapproachmusicsegmentation,
      title={A Knowledge-Driven Approach to Music Segmentation, Music Source Separation and Cinematic Audio Source Separation}, 
      author={Chun-Wei Ho and Sabato Marco Siniscalchi and Kai Li and Chin-Hui Lee},
      year={2026},
      eprint={2602.21476},
      archivePrefix={arXiv},
      primaryClass={eess.AS},
      url={https://arxiv.org/abs/2602.21476}, 
}

\end{document}